\documentclass[12pt]{article}
\usepackage{amsmath}
\usepackage{amssymb}
\usepackage{graphicx}
\usepackage{url}
\usepackage{hyperref}
\usepackage{natbib}
\usepackage{setspace}
\usepackage[margin=1in]{geometry}
\title{The Risk-Adjusted Intelligence Dividend: A Quantitative Framework for Measuring AI Return on Investment Integrating ISO 42001 and Regulatory Exposure}
\author{Prof. Hernan Huwyler, MBA CPA}
\date{November 26, 2025}

\begin{document}

\maketitle

\begin{abstract}
Organizations investing in artificial intelligence face a fundamental challenge: traditional return on investment calculations fail to capture the dual nature of AI implementations, which simultaneously reduce certain operational risks while introducing novel exposures related to algorithmic malfunction, adversarial attacks, and regulatory liability. This research presents a comprehensive financial framework for quantifying AI project returns that explicitly integrates changes in organizational risk profiles. The methodology addresses a critical gap in current practice where investment decisions rely on optimistic benefit projections without accounting for the probabilistic costs of AI-specific threats including model drift, bias-related litigation, and compliance failures under emerging regulations such as the European Union Artificial Intelligence Act and ISO/IEC 42001. Drawing on established risk quantification methods, including annual loss expectancy calculations and Monte Carlo simulation techniques, this framework enables practitioners to compute net benefits that incorporate both productivity gains and the delta between pre-implementation and post-implementation risk exposures. The analysis demonstrates that accurate AI investment evaluation requires explicit modeling of control effectiveness, reserve requirements for algorithmic failures, and the ongoing operational costs of maintaining model performance. Practical implications include specific guidance for establishing governance structures, conducting phased validations, and integrating risk-adjusted metrics into capital allocation decisions, ultimately enabling evidence-based AI portfolio management that satisfies both fiduciary responsibilities and regulatory mandates.
\end{abstract}

\textbf{Keywords:} risk-adjusted ROI, artificial intelligence investment, AI governance, quantitative risk assessment, total cost of ownership, annual loss expectancy, compliance frameworks

\textbf{JEL Codes:} G31; M15; O32; D81; G32

\section{Introduction}

The acceleration of artificial intelligence adoption across industries has created an urgent need for financial frameworks capable of accurately quantifying investment returns. Industry research, such as the Deloitte’s 2025 AI ROI survey, indicates that most organizations require two to four years to achieve payback on AI investments, while median returns hover around ten percent with approximately one-third of implementations generating limited or negligible gains \citep{stange2025}. This pattern suggests systematic failures in how organizations evaluate and manage AI investments rather than fundamental limitations of the technology itself.

Current approaches to AI return on investment suffer from a critical methodological weakness: they treat AI implementations as conventional technology projects amenable to standard cost-benefit analysis. This conceptualization ignores the distinctive characteristic of AI systems, namely their capacity to fundamentally alter an organization's risk profile in both directions. Automated decision-making can eliminate human error in routine processes while simultaneously introducing algorithmic vulnerabilities that create entirely new categories of potential loss. A fraud detection system may reduce financial losses from criminal activity while exposing the organization to discrimination lawsuits if the underlying model exhibits demographic bias.

The regulatory landscape has intensified this challenge. The European Union Artificial Intelligence Act establishes a risk-based classification system with penalties reaching thirty-five million euros or seven percent of global annual turnover for violations involving prohibited AI practices \citep{euparl2024}. ISO/IEC 42001:2023 requires formal risk assessment processes throughout the AI lifecycle \citep{iso42001}. The National Institute of Standards and Technology Artificial Intelligence Risk Management Framework requires organizations to continuously measure and manage AI-specific risks \citep{nist2023}. These requirements transform risk quantification from a discretionary analytical exercise into a compliance obligation with direct financial consequences.

This research addresses the operational gap between governance mandates and practical implementation by presenting a risk-adjusted return on investment framework explicitly designed for AI projects. The methodology draws on established quantitative risk assessment techniques, adapting annual loss expectancy calculations and probability distribution modeling to the unique characteristics of algorithmic systems. The framework enables organizations to answer fundamental economic questions: What contingency reserves should be established for AI risks? What is the return on investment for specific control implementations? How should warranty and insurance provisions be structured for AI-enabled products and services?

The following sections establish the current state of knowledge regarding AI investment evaluation and risk frameworks, identify specific gaps in existing methodologies, present a comprehensive solution architecture, and analyze the governance implications of risk-adjusted AI financial modeling.

\section{Research and Analysis}

\subsection{Current State of Knowledge}

The evaluation of AI investments occurs within a complex ecosystem of financial methodologies, governance frameworks, and regulatory requirements. Understanding this landscape is essential for identifying where current approaches fail and how improvements can address practitioner needs.

\subsubsection{Financial Evaluation Methodologies}

Traditional capital budgeting techniques provide the foundation for AI investment analysis. Net present value calculations discount future cash flows to establish current worth, internal rate of return identifies the discount rate at which net present value equals zero, and payback period measures the time required to recover initial outlays. These methods assume cash flows can be estimated with reasonable accuracy and that risk can be captured through discount rate adjustments or scenario analysis.

Applied to AI projects, these techniques require translation of algorithmic capabilities into financial metrics. Productivity benefits derive from automation of manual processes, where freed labor hours multiplied by fully-loaded employee costs generate quantifiable savings. Revenue enhancements emerge from AI-enabled products, personalized recommendations, predictive lead scoring, and dynamic pricing mechanisms. The challenge lies not in the calculation mechanics but in the validity of the underlying assumptions and the completeness of cost and risk identification.

Industry benchmarks reveal significant variability in AI investment outcomes. Research examining finance function AI implementations found median returns of approximately ten percent, with top performers achieving substantially higher results through disciplined value-focused approaches \citep{stange2025}. The variance between high and low performers suggests that methodology and execution quality significantly influence outcomes independent of the underlying technology capabilities.

Despite the availability of these frameworks, practical application is hindered by significant operational gaps. The primary challenge is the identification and quantification of "Hidden Technical Debt." Organizations frequently calculate the Return on Investment (ROI) based solely on the initial development and deployment costs (CapEx), ignoring the substantial continuous costs (OpEx) required to keep a model viable.

\begin{itemize}
\item \textbf{The TCO Blind Spot:} Traditional models account for hardware and software licenses but fail to budget for the "continuous" nature of AI. This includes the costs of data pipeline maintenance, concept drift detection, retraining cycles, and the premium salaries required for MLOps talent. Research from the IBM Institute for Business Value indicates that technical debt in machine learning systems can erode projected ROI by 18-29\% if left unserviced.
\item \textbf{The Attribution Problem:} Without rigorous baselining, organizations fall prey to "phantom benefits." If the pre-AI process performance is not documented through time-motion studies or process mining, any post-deployment improvement is attributed to AI, even if it resulted from exogenous market factors. This lack of causal attribution leads to inflated ROI claims that dissolve under audit.
\item \textbf{Unpriced Risk Exposure:} Perhaps the most critical gap is the failure to monetize risk. A project may appear profitable because the cost of potential regulatory sanctions, reputation damage from bias, or liability from automated errors is excluded from the balance sheet. Conversely, projects that significantly reduce fraud or safety incidents are undervalued because "risk avoidance" is not calculated as a revenue stream.
\end{itemize}

\subsubsection{Governance and Compliance Frameworks}

ISO/IEC 42001:2023 establishes the first international standard for AI management systems, requiring organizations to define objectives and success criteria, conduct AI-specific risk assessments, implement controls proportionate to identified risks, and maintain continuous monitoring and improvement processes \citep{iso42001}. Clause 6.1 requires systematic identification and evaluation of AI risks, while Clause 9 requires performance measurement against defined indicators. The standard explicitly connects governance activities to measurable outcomes, creating a foundation for integrating compliance requirements with financial evaluation.

The National Institute of Standards and Technology Artificial Intelligence Risk Management Framework provides complementary guidance through its four core functions: Govern, Map, Measure, and Manage \citep{nist2023}. The Measure function directly supports financial modeling by requiring organizations to track metrics related to AI system trustworthiness, including accuracy, fairness, and security characteristics. Importantly, the framework acknowledges that AI systems may require more frequent maintenance than traditional software due to data drift and concept drift, implying ongoing operational costs that must factor into total cost of ownership calculations \citep{nist2023}.

The European Union Artificial Intelligence Act introduces legally binding requirements with substantial financial penalties. High-risk AI systems must maintain documented risk management systems covering the complete lifecycle, implement data governance practices ensuring training data quality, provide transparency through technical documentation and user instructions, and enable human oversight of automated decisions \citep{euparl2024}. Non-compliance penalties scale with violation severity: up to thirty-five million euros or seven percent of turnover for prohibited practices, up to fifteen million euros or three percent for high-risk system violations, and up to seven point five million euros or one percent for information provision failures \citep{euparl2024}. These penalty structures create quantifiable downside scenarios that must inform investment analysis.

\subsubsection{Risk Quantification Methods}

Established approaches to risk quantification provide methodological foundations applicable to AI contexts. The Factor Analysis of Information Risk framework models risk as a function of loss event frequency and probable loss magnitude, enabling monetary quantification of potential adverse outcomes. Annual loss expectancy calculations multiply single loss expectancy by annualized rate of occurrence to establish expected yearly costs from specific risk scenarios.

Monte Carlo simulation techniques enable modeling of complex risk scenarios where multiple uncertain variables interact. By running thousands of iterations with sampled parameter values, these methods generate probability distributions of potential outcomes rather than single point estimates \citep{huwyler2025a}. This approach proves particularly valuable for AI risks where historical data may be limited but expert judgment can inform probability distributions for novel threat vectors.

Research in quantitative risk assessment demonstrates that computational frameworks can effectively model uncertainty across diverse scenarios, providing confidence intervals that support reserve setting and capital allocation decisions \citep{huwyler2025b}. Applied to AI investments, these techniques enable organizations to move beyond simplistic best-case and worst-case scenarios toward probabilistically weighted financial projections.

\subsection{Practical Gaps and Challenges}

Despite the availability of financial methodologies and governance frameworks, significant gaps impede effective AI investment evaluation in practice.

\subsubsection{Incomplete Cost Identification}

Organizations consistently underestimate total cost of ownership for AI systems. Initial focus on hardware acquisition, software licensing, and development team costs obscures substantial ongoing expenditures. Cloud compute and storage consumption fees accumulate based on inference volumes that may be difficult to forecast. Data pipeline maintenance requires continuous investment as source systems evolve. Model retraining cycles address performance degradation but consume engineering resources and compute capacity. Compliance monitoring and audit documentation impose overhead that scales with regulatory scrutiny.

Research demonstrates that machine learning systems accumulate significant technical debt across multiple dimensions, including configuration complexity, data dependencies, and system-level anti-patterns, that can substantially erode projected returns over time \citep{sculley2015}. This erosion occurs gradually as organizations defer maintenance activities, accumulate workarounds, and allow system quality to degrade. Traditional financial models rarely incorporate explicit provisions for technical debt remediation, leading to systematic overestimation of net benefits.

\subsubsection{Risk Profile Changes Unquantified}

The most significant gap in current practice involves the failure to quantify how AI implementations change organizational risk exposures. AI systems do not simply add capabilities; they fundamentally restructure the risk landscape by eliminating certain threats while introducing others.

Consider a loan underwriting system that replaces manual credit analysis with algorithmic decisioning. The AI implementation may reduce errors from human fatigue, eliminate inconsistency across different analysts, and accelerate processing time. These benefits translate into quantifiable productivity gains and potentially reduced credit losses from improved prediction accuracy. However, the same system introduces risks absent from the manual process: algorithmic bias that may violate fair lending regulations, model drift as economic conditions change borrower behavior, adversarial attacks that manipulate input features to obtain unwarranted approvals, and opacity that complicates regulatory examination responses.

Accurate investment evaluation requires calculating the risk delta: the difference between annual loss expectancy under the AI-enabled process and annual loss expectancy under the existing process. A positive delta indicates net risk reduction that adds value; a negative delta represents increased exposure that should be treated as an ongoing cost. Current methodologies acknowledge this conceptually but rarely operationalize it through explicit quantification.

\subsubsection{Benefit Attribution Challenges}

Isolating AI's contribution to observed improvements proves methodologically difficult. Multiple factors influence business outcomes simultaneously, and post-implementation improvements may reflect general market conditions, concurrent process changes, or regression to the mean rather than AI effectiveness. Without rigorous attribution methods, organizations cannot validate their investment assumptions or learn from implementation experience.

The challenge intensifies for revenue enhancement use cases where AI influences customer behavior through complex causal pathways. A recommendation engine may increase conversion rates, but the observed uplift could partially reflect improved website design, seasonal demand patterns, or marketing campaign timing. Cannibalization effects further complicate attribution when AI-driven sales in one channel displace revenue from others.

\subsection{Proposed Solution: Risk-Adjusted AI Investment Framework}

Addressing identified gaps requires a comprehensive framework that integrates financial modeling with risk quantification and governance requirements. The following methodology enables practitioners to develop investment analyses that withstand scrutiny from financial, technical, and regulatory perspectives.

\subsubsection{Framework Architecture}

The risk-adjusted return on investment calculation modifies traditional formulas to incorporate explicit risk components:

\begin{equation}
\begin{aligned}
Risk\text{-}Adjusted\ ROI = & \ Gross\ Benefits + Risk\ Reduction\ Benefits \\
& - Risk\ Increase\ Costs - TCO
\end{aligned}
\end{equation}

This formulation recognizes four distinct value and cost categories that must be independently quantified:

Gross benefits encompass productivity gains from automation and optimization plus revenue enhancements from AI-enabled products and services. These represent the traditional focus of AI business cases.

Risk reduction benefits capture the monetary value of decreased loss exposure in areas where AI improves upon existing processes. Reduced fraud losses, fewer compliance violations, and lower error remediation costs contribute to this category.

Risk increase costs represent the expected losses from AI-specific threats including model failures, adversarial attacks, bias-related liability, and regulatory non-compliance. These costs often receive inadequate attention in conventional analyses.

Total cost of ownership includes complete lifecycle expenditures across capital and operating categories: infrastructure, licensing, personnel, data management, monitoring, retraining, compliance, and security.

\subsubsection{Value Quantification Methodology}

Productivity benefit calculation requires systematic process analysis before AI implementation. Process mining tools identify automatable workflows and establish baseline metrics for manual task duration, error rates, and throughput volumes. Time-motion studies over two to four week periods capture actual labor consumption patterns that may differ significantly from management estimates.

The conversion from operational metrics to financial impact follows a structured approach. Freed hours multiply by fully-loaded employee costs including salary, benefits, facilities, and management overhead. Error reduction translates into avoided rework costs, warranty claims, and compliance violation penalties using historical cost data. Quality improvements reduce customer complaints, returns, and support contacts, each with quantifiable cost implications.

Revenue enhancement quantification demands controlled experimentation. A/B testing with randomly assigned treatment and control groups isolates AI's incremental contribution to conversion rates, average order values, or customer lifetime values. Synthetic control methods provide alternatives when randomization proves infeasible. The critical discipline involves using actual measured uplift rather than vendor projections or optimistic assumptions.

Importantly, benefit projections should incorporate error margins of twenty to thirty percent during early implementation phases when limited performance data exists. As operational experience accumulates, these margins can narrow based on observed variance between projections and actuals.

\subsubsection{Risk Delta Calculation}

Risk profile changes require explicit modeling of both risk reduction and risk introduction effects.

For each AI-related risk scenario, the analysis establishes single loss expectancy based on the financial impact should the event occur, considering direct costs, remediation expenses, regulatory penalties, and reputational damage. Annual rate of occurrence derives from threat modeling, industry benchmarks, and expert judgment when historical data proves unavailable. The product yields annual loss expectancy for that scenario.

Risk reduction scenarios represent threats that AI mitigates more effectively than existing processes. A fraud detection system may reduce annual loss expectancy from criminal activity by identifying patterns invisible to manual review. The difference between pre-AI and post-AI annual loss expectancy for these scenarios represents value created.

Risk introduction scenarios capture AI-specific threats absent from current processes. Model inversion attacks, training data poisoning, algorithmic bias, concept drift, and hallucinated outputs each require annual loss expectancy estimation. For high-risk AI systems under the European Union Artificial Intelligence Act, scenarios must include regulatory penalty exposure scaled by violation probability.

The aggregate risk delta sums individual scenario calculations:

\begin{equation}
\text{Risk Delta} = \sum (\text{ALE}_{\text{current}} - \text{ALE}_{\text{AI}}) \quad \text{for all identified scenarios}
\end{equation}

Positive values indicate net risk reduction; negative values require treatment as ongoing costs that reduce net benefits.

\subsubsection{Total Cost of Ownership Components}

Comprehensive cost identification addresses capital and operating categories across the complete system lifecycle.

Capital expenditures include infrastructure acquisition, initial software licensing, training data procurement, and development team mobilization. These costs require amortization across expected useful life to enable meaningful comparison with ongoing benefits.

Operating expenditures span cloud compute consumption, data pipeline maintenance, model monitoring and observability tooling, retraining cycles, personnel costs for data scientists and MLOps engineers, compliance audit and documentation overhead, security testing and vulnerability remediation, and insurance premium increases reflecting AI risk exposures.

Research suggests that organizations should budget for model maintenance at fifteen to twenty-five percent of initial development costs annually, with talent premiums for specialized machine learning engineers commanding thirty to fifty percent above comparable software engineering compensation \citep{gartner2023}. Some insurers may charge higher premiums for AI‑intensive environments due to the expanded attack surface and emerging risk uncertainty.

Reserve requirements represent an often-neglected cost category. Best practice suggests maintaining risk reserves at ten to fifteen percent of AI operational budgets to address unmodeled issues, emerging threats, and unexpected performance degradation. These reserves should factor into net benefit calculations as allocated but unavailable capital.

\subsection{Technical and Governance Analysis}

The proposed framework creates value through multiple mechanisms while imposing implementation requirements that organizations must realistically assess.

\subsubsection{Alignment with Regulatory Requirements}

The methodology directly supports compliance with major governance frameworks. ISO/IEC 42001 requires organizations to define AI system objectives, conduct risk assessments, and monitor performance against established criteria \citep{iso42001}. The framework's explicit benefit metrics, risk scenario analysis, and ongoing monitoring provisions satisfy these requirements while generating documentation suitable for audit purposes.

The National Institute of Standards and Technology Artificial Intelligence Risk Management Framework's Map and Measure functions find direct expression in the risk delta calculation process \citep{nist2023}. Systematic identification of risk scenarios operationalizes the mapping requirement, while annual loss expectancy calculations provide the quantitative measures the framework demands.

European Union Artificial Intelligence Act compliance requires documented risk management systems for high-risk applications \citep{euparl2024}. The framework generates the risk registers, control documentation, and performance records that regulators will examine. More importantly, explicit penalty exposure calculations ensure that compliance costs receive appropriate weight in investment decisions rather than treatment as afterthoughts.

\subsubsection{Decision Quality Improvement}

Beyond compliance, the framework enhances investment decision quality through several mechanisms.

Explicit uncertainty modeling through probability distributions and confidence intervals prevents false precision that leads to overcommitment. Presenting returns as ranges with associated probabilities enables more nuanced capital allocation than single-point estimates that obscure underlying assumptions.

Causal attribution requirements imposed by the benefit quantification methodology ensure that claimed gains reflect actual AI contribution rather than concurrent improvements or measurement artifacts. Organizations learn from experience when rigorous attribution validates or invalidates their assumptions.

Continuous recalculation through quarterly return on investment reviews with actual performance data enables course correction when implementations deviate from projections. This iteration discipline converts static business cases into dynamic management tools.

\subsubsection{Implementation Challenges}

Adopting the framework requires organizational capabilities that many enterprises currently lack.

Data infrastructure must support both baseline measurement and ongoing performance tracking. Process mining tools, financial systems, and operational databases require integration to enable the calculations the framework demands. Organizations with fragmented data landscapes will find implementation significantly more difficult.

Analytical expertise spanning finance, statistics, and machine learning must be available to develop and validate models. The interdisciplinary nature of risk-adjusted AI evaluation exceeds the capabilities of traditional finance or technology teams operating independently.

Governance processes must accommodate the framework's requirements for controlled experimentation, periodic recalculation, and documented assumption validation. Organizations accustomed to one-time business case approval followed by minimal oversight will need to establish new review rhythms and accountability structures.

\section{Discussion}

\subsection{Interpretation of Framework Efficacy}

The risk-adjusted return on investment framework addresses fundamental weaknesses in how organizations evaluate AI investments. By requiring explicit quantification of risk profile changes, the methodology prevents the systematic optimism that industry data suggests characterizes most AI business cases. The observation that median returns hover around ten percent while top performers achieve substantially better results \citep{stange2025} indicates that methodology quality significantly influences outcomes. Organizations adopting rigorous frameworks can reasonably expect to shift their performance toward the upper distribution.

The framework's integration of compliance requirements with financial analysis resolves the artificial separation between governance and value creation that characterizes many organizations. Rather than treating regulatory compliance as an overhead cost imposed on otherwise valuable projects, the methodology recognizes compliance as risk mitigation that contributes to sustainable returns. This reframing proves particularly important as penalty structures under the European Union Artificial Intelligence Act create material downside exposures that must inform capital allocation \citep{euparl2024}.

\subsection{Practical Implications and Recommendations}

Organizations implementing the framework should prioritize several immediate actions.

Establish governance structures before initiating AI investments. An AI investment committee with representation from finance, risk, technology, and legal functions provides the cross-functional perspective that accurate evaluation requires. Executive sponsorship ensures that rigorous methodology receives organizational support rather than dismissal as bureaucratic obstruction.

Develop baseline measurement capabilities for candidate AI use cases. Without pre-implementation performance data, benefit attribution becomes impossible and the framework's value diminishes substantially. Investment in process mining tools and time-motion study protocols pays dividends across multiple projects.

Mandate presentation of returns as probability distributions rather than point estimates. The discipline of articulating confidence intervals forces explicit acknowledgment of uncertainty and prevents overcommitment based on best-case assumptions. A requirement that business cases include tenth percentile, fiftieth percentile, and ninetieth percentile projections provides decision-makers with more actionable information.

Integrate risk reserve requirements into capital allocation processes. Treating reserves as committed costs rather than optional provisions ensures that portfolios remain resilient to adverse scenarios. The ten to fifteen percent reserve guideline provides a reasonable starting point pending organization-specific calibration.

\subsection{Limitations}

The framework's effectiveness depends on data availability and analytical capability that vary across organizations. Enterprises with mature data infrastructure and quantitative talent will find implementation substantially easier than those requiring foundational investments before methodology adoption.

The methodology assumes that meaningful probability distributions can be developed for AI risk scenarios, which proves challenging for novel threats lacking historical precedent. Organizations should acknowledge the inherent uncertainty in forward-looking risk quantification while recognizing that explicit estimation with acknowledged limitations improves upon implicit assumptions embedded in conventional analysis.

The framework does not address broader strategic considerations such as competitive positioning, organizational learning, and option value from AI capability development. These factors may justify investments that pure financial analysis would reject, suggesting that the methodology should inform rather than dictate strategic decisions.

\subsection{Overall Contribution}

This research contributes to the emerging discipline of AI financial governance by providing an actionable methodology that bridges theoretical frameworks and practical implementation. The explicit integration of risk profile changes into return calculations addresses the most significant gap in current practice while the alignment with regulatory requirements ensures relevance as compliance obligations intensify.

\section{Conclusions}

\subsection{Summary of Key Findings}

Effective AI investment evaluation requires explicit quantification of risk profile changes alongside traditional cost and benefit analysis. Organizations implementing AI systems simultaneously reduce certain operational risks while introducing novel exposures from algorithmic malfunction, adversarial attacks, bias-related liability, and regulatory non-compliance. Accurate return on investment calculations must incorporate the risk delta representing net change in annual loss expectancy between pre-implementation and post-implementation states.

Total cost of ownership for AI systems extends substantially beyond initial capital expenditure to include ongoing data management, model maintenance, compliance overhead, and risk reserves. Industry research suggesting a high technical debt erosion \citep{sculley2015} and implementation timelines of two to four years for payback achievement indicate that conventional cost estimation systematically underweights these factors.

Regulatory frameworks including ISO/IEC 42001, the National Institute of Standards and Technology Artificial Intelligence Risk Management Framework, and the European Union Artificial Intelligence Act mandate documented risk management processes that the proposed framework directly supports \citep{iso42001, nist2023, euparl2024}. Penalty structures reaching seven percent of global turnover for serious violations create material downside exposures that must inform investment analysis.

\subsection{Contribution to the Field}

This research advances the practice of AI financial governance by providing a comprehensive methodology that integrates financial analysis with risk quantification and compliance requirements. The framework enables practitioners to answer fundamental economic questions that current approaches leave unaddressed: appropriate reserve levels for AI risks, return on investment for specific controls, and optimal structuring of warranties and insurance provisions.

The explicit risk delta calculation addresses the most significant gap in current practice where investment decisions proceed based on optimistic benefit projections without systematic consideration of introduced risks. By requiring quantification of both risk reduction and risk introduction effects, the methodology ensures that capital allocation reflects complete value and cost profiles.

\subsection{Recommendations for Practitioners}

Organizations should implement several immediate actions to improve AI investment evaluation quality.

Form cross-functional AI investment committees with finance, risk, technology, and legal representation to ensure comprehensive evaluation perspectives.

Establish baseline measurement protocols for candidate AI use cases including process mining, time-motion studies, and historical error analysis before project approval.

Mandate probability distribution presentation of projected returns with explicit tenth, fiftieth, and ninetieth percentile scenarios rather than single-point estimates.

Incorporate risk reserve requirements of ten to fifteen percent of AI operational budgets into capital allocation and project financial models.

Conduct quarterly return on investment recalculations using actual performance data to enable course correction and organizational learning.

Update vendor risk assessments and procurement processes to require ISO 42001 certification, documented risk management practices, and performance-based pricing with service level penalties.

\subsection{Future Research Directions}

Several areas warrant additional investigation to advance the discipline of AI financial governance.

Development of industry-specific benchmarks for AI risk scenario annual loss expectancy would enable more accurate modeling by organizations lacking extensive internal historical data. Collaborative efforts to anonymize and aggregate incident data could support such benchmarking while protecting competitive information.

Investigation of relationships between governance maturity and AI investment outcomes could validate the proposed framework's impact and identify additional success factors. Longitudinal studies tracking organizations through AI implementation cycles would provide particularly valuable evidence.

Analysis of emerging accounting standards for software and AI assets, including their implications for financial reporting of AI investments, would help practitioners navigate evolving disclosure requirements. The interaction between capitalization rules and risk-adjusted return calculations merits specific attention.

Research examining optimal reserve levels for different AI application categories and risk profiles would refine current guidance based on empirical observation rather than expert judgment alone. As the installed base of AI systems grows, sufficient data should accumulate to support actuarial analysis of AI-related losses.

\section*{References}

\bibliographystyle{apalike}
\bibliography{references}

\section*{Conflicts of Interest}
The author declares no conflicts of interest.

\section*{Funding}
This research received no external funding.

\section*{Version}
1.0

\section*{Date}
November 26, 2025

\section*{Author Biography}
Prof. Hernan Huwyler, MBA CPA serves as the Academic Director for Compliance, Risk Management, and AI programs at IE Business School and IE Law School. Concurrently, he is a Senior Manager at Capgemini Invent's Applied AI Lab (Nordics), leading enterprise-wide AI governance, risk modeling, and financial control initiatives. A recognized expert in the intersection of AI investment evaluation and regulatory compliance, Prof. Huwyler specializes in operationalizing risk-adjusted financial frameworks aligned with ISO 42001, the NIST AI Risk Management Framework, and the European Union Artificial Intelligence Act. His research focuses on quantitative methodologies for transforming abstract AI capabilities into auditable financial metrics that satisfy both fiduciary responsibilities and emerging governance mandates.

\end{document}